\pdfoutput=1

\documentclass[11pt,a4paper]{article}

\usepackage[T1]{fontenc}
\IfFileExists{newtxtext.sty}{\usepackage{newtxtext,newtxmath}}{\usepackage{mathptmx}}
\usepackage[margin=2.5cm]{geometry}
\usepackage{graphicx}
\usepackage{amsmath}
\usepackage[round]{natbib}
\usepackage[hidelinks]{hyperref}
\usepackage{url}
\usepackage[section,above]{placeins}
\providecommand{\Description}[2][]{}

\title{\textbf {GOWC: an open tool for observing-conditions prediction and observation
planning, tested against site monitors and an exposure-time calculator}}

\author{Ahzam Ahmed\thanks{E-mail: \texttt{ahzamahd641@gmail.com};
ORCID: \href{https://orcid.org/0009-0007-3864-0934}{0009-0007-3864-0934}}\\[2pt]
\normalsize Independent researcher, United Kingdom}

\date{}

\begin{document}
\maketitle

\begin{abstract}
GOWC (Global Observing-conditions Weather Catalogue) is an open web application that takes the user from a target to an exposure plan across 2,642 catalogued observatory sites. It computes seeing, airmass, extinction, precipitable water vapour, sky brightness and signal-to-noise from live Open-Meteo meteorology, including pressure-level profiles, and tests most of these outputs against an independent reference. The photometric SNR calculator, run with GOWC's own catalogue entries,
reproduces the ING \textsc{signal} exposure-time calculator to a median $1.9$~per~cent over a
16-point magnitude--airmass grid on two instruments. Precipitable water vapour tracks
the Paranal LHATPRO radiometer closely ($\rho=+0.95$, 606 nights) with a $+43$~per~cent wet
bias. Seeing is computed from the Tatarski $C_n^2$ chain integrated over the pressure
levels above the site and calibrated to the ESO MASS free-atmosphere median at
Paranal; matching it requires an effective Tatarski coefficient of $5.7$, nearly twice the upper end of the published range, which quantifies how much turbulence four coarse pressure levels
fail to resolve. Against $\sim$20 months of ESO DIMM data the seeing forecast has a
weak but significant night-to-night ranking skill ($\rho=+0.24$, 575 nights, at
Paranal; $+0.16$ at La Silla). Surface wind alone does as well or better ($+0.27$, $+0.41$) and
persistence does better ($+0.42$, $+0.47$), but the free-atmosphere term tracks MASS
($\rho=+0.35$) and carries information that surface wind does not (partial
$\rho=+0.28$, $+0.40$). At held-out sites the absolute seeing
scale is no better than a constant. GOWC therefore treats forecast seeing as
indicative, shows measured values wherever a public site monitor exists, and states
which outputs are measured, calibrated or unvalidated.
\end{abstract}



\section{Introduction}

Planning a night at an optical telescope involves several steps. The observer
first confirms that the target reaches a usable altitude during astronomical
darkness, then estimates the conditions for that night (seeing, transparency and
sky brightness with the Moon up), and finally sets an exposure time that reaches
the target signal-to-noise ratio (SNR). Each step depends on the one before it,
but the tools that serve them are usually separate. Visibility comes from
planetarium software, conditions from a weather page or forecast, and the SNR
from an instrument-specific exposure-time calculator (ETC).

Dedicated tools already address parts of this chain. Astroplan computes target
visibility \citep{astroplan}, the ESO ETCs and \textsc{skycalc} \citep{skycalc} model sky
brightness and exposure times, and mesoscale turbulence forecasts
such as MOSE predict seeing at individual sites \citep{masciadri}. Large
facilities also run their own ETCs and site monitors. Most catalogued
observatories, however, including many small professional, university and amateur
stations, have neither. An observer at such a site, or one comparing several sites
for a single target, has to combine general-purpose tools that were not designed
to work together.

GOWC (Global Observing-conditions Weather Catalogue) brings this workflow into one
open web application covering $2{,}642$ catalogued observatory sites. Its design
priority is that every reported number can be traced to a published physical
model or an explicit calibration, with its assumptions stated. This paper
describes those models, tests them against site monitors and a professional ETC,
and identifies where the method breaks down. The results support exposure-time
planning, airmass and dark-time scheduling, and a water-vapour forecast with a
known bias. Forecast seeing is indicative only.

This paper makes four contributions. First, it sets out a reproducible forward
planning chain covering visibility, seeing, extinction, precipitable water vapour,
sky brightness and photometric SNR, driven by one shared physics module across a
catalogue of $2{,}642$ sites rather than a single facility. Second, it benchmarks
the SNR calculation against a professional ETC across a grid in magnitude and
airmass, using the instrument constants an observer actually gets. Third, it
measures the night-to-night skill of the seeing forecast over a 20-month hindcast,
in which the model is run on archived weather and compared with ESO Astronomical
Site Monitor measurements. It compares that skill with two baselines an observer
already has, persistence and surface wind, and reports what the calibration
implies about the input data. Fourth, it states for each quantity whether the
output is backed by measurement, calibrated, or passed through without validation.

Section~\ref{sec:system} describes the system, Section~\ref{sec:models} the
physical models and Section~\ref{sec:validation} their validation.
Section~\ref{sec:discussion} discusses the limitations and
Section~\ref{sec:future} outlines future work.

\section{System Overview}\label{sec:system}
\label{sec:overview}

GOWC is a Python application with a database back end, a web dashboard and a
REST API. The observatory list is built from the IAU
Minor Planet Center observatory-code list    \citep[retrieved 2026-10-16;][]{mpc_obscodes}, with several flagship research
   facilities added by hand, giving $2{,}642$ catalogued sites. The MPC codes include
many small and amateur stations, so this is a count of catalogued observatory sites
rather than of major professional facilities. Each site's geodetic latitude is
recovered from the MPC parallax constants $(\rho\cos\phi',\rho\sin\phi')$ by the
standard iterative geocentric-to-geodetic conversion on the WGS84 ellipsoid, and its
elevation is taken from a digital-elevation lookup at those coordinates. (The
geocentric latitude $\phi'=\operatorname{atan2}(\rho\sin\phi',\rho\cos\phi')$ lies up
to $0.19^\circ$, $\sim$21\,km, equatorward of the true site, enough to move an
elevation lookup by hundreds of metres in mountainous terrain, so the conversion
matters.) About $1.4$~per~cent of MPC entries carry
only coarse coordinates (longitude to $\le$0.01$^\circ$ or parallax constants to
$\le$3 decimals); for those the elevation, and hence the altitude-dependent seeing
and extinction, remains uncertain.

\begin{figure}[htbp]
\centering
\includegraphics[width=\textwidth]{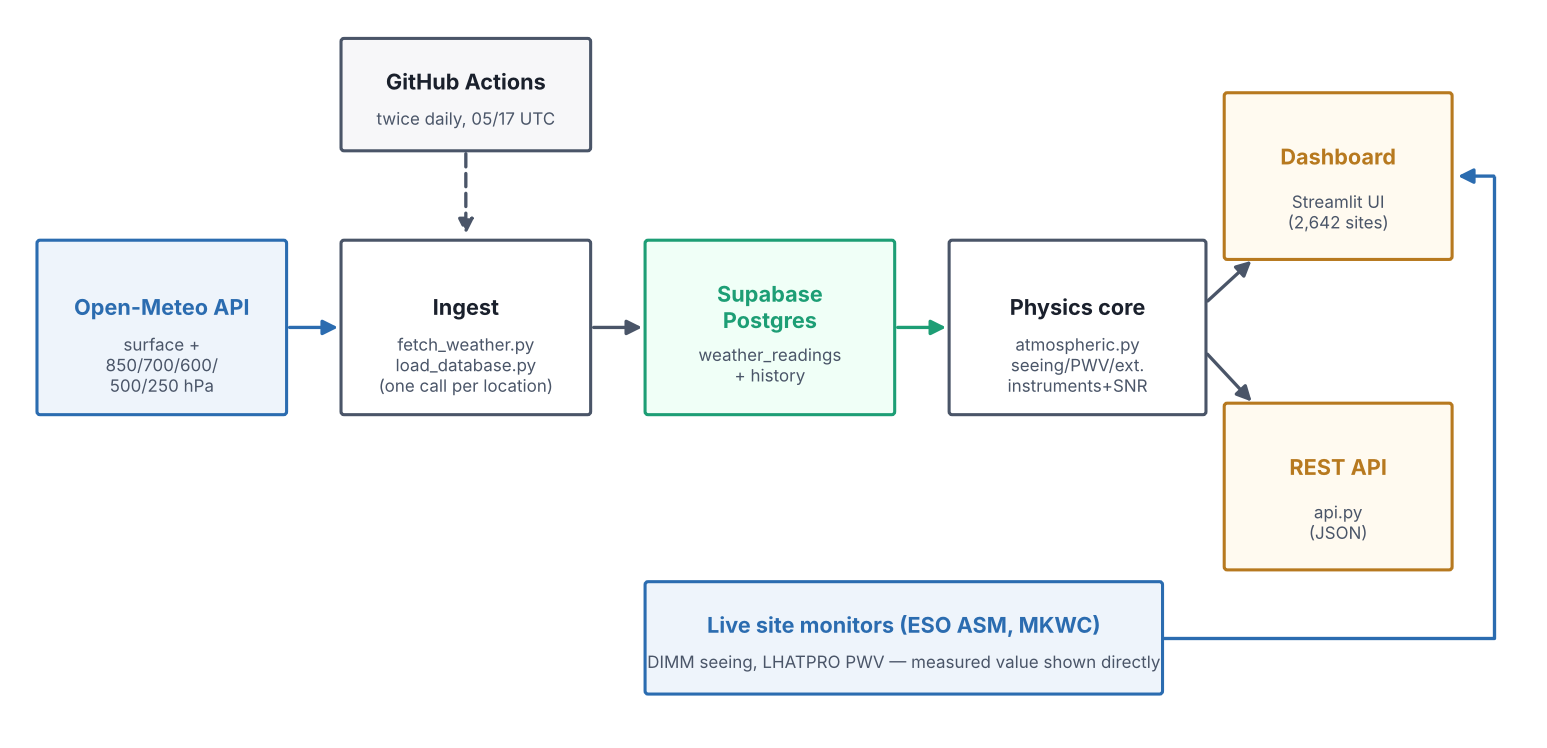}
\caption{GOWC data flow. A scheduled GitHub Actions job (twice daily) pulls
surface and pressure-level fields from Open-Meteo, one request per distinct
location, and loads them into a Supabase Postgres store; a shared physics core
(\texttt{atmospheric.py} plus the instrument/SNR catalogue) converts them into
observing quantities, served by an interactive Streamlit dashboard and a REST
API. Where a public site monitor publishes live data (ESO ASM DIMM and LHATPRO
at Paranal, Maunakea Weather Center DIMM), the dashboard shows the
\emph{measured} value directly, bypassing the forecast model.}
\Description[Flow diagram of the GOWC system.]{Flow diagram. Open-Meteo weather data pass through an ingest step, run twice daily by GitHub Actions, into a Postgres database and then a shared physics core, which feeds both an interactive dashboard and a REST API. A separate box for live site monitors (ESO and Maunakea DIMM seeing, LHATPRO water vapour) connects directly to the dashboard, bypassing the forecast model.}%
\label{fig:arch}
\end{figure}

Fig.~\ref{fig:arch} shows the data flow. A scheduled job (GitHub Actions, twice daily at 05 and 17\,UTC) fetches weather for
every observatory from Open-Meteo, one request per distinct location shared by
co-located sites. Each request returns surface fields (cloud, humidity, wind,
temperature, dew point, pressure, precipitation), pressure-level profiles
(temperature, geopotential height and wind at 850, 700, 600 and 500\,hPa, wind at
250\,hPa for the jet stream, and relative humidity at 850 and 500\,hPa), and the
model total-column water vapour. All
winds are requested in m\,s$^{-1}$. Each fetch stores the fields for the forecast hour
nearest the fetch time, so the application reports conditions at that hour rather
than a night-integrated forecast; at 17\,UTC that hour is daytime in Chile. A shared
physics module converts the stored fields into observing quantities, served by two
front-ends, an interactive dashboard and a REST API, both of which pass the full
pressure-level profile to the same functions. A guided planning workflow runs on top
of the physics engine: it handles target and site selection, computes visibility and
the best window, applies the conditions, solves for exposure and SNR, and produces an
exportable observing plan.

\subsection{Scope: the forward half of the pipeline}
The standard photometric
workflow has two halves. The \emph{forward} (predictive) half turns a target
magnitude into expected detector counts and a signal-to-noise ratio, through the
chain, magnitude $\to$ flux $\to$ atmospheric extinction $\to$ collected photons
$\to$ detected electrons $\to$ SNR. The \emph{reverse} (reduction) half recovers a calibrated magnitude
from a real frame: centroiding and PSF fitting, aperture and annulus sky
estimation, relative photometry against comparison stars, and zero-point
calibration against standard stars. GOWC implements the forward half, including the Vega/UBVRI system and the
Bouguer extinction law. The engine also supports the secondary,
colour-dependent extinction term and the instrumental-to-standard colour term,
although the current interface does not yet pass a target colour
(Section~\ref{sec:airmass}). It also predicts the magnitude uncertainty
$\sigma_m = 1.0857/\mathrm{SNR}$ that the reverse half would ultimately measure.
It does not perform the reduction steps, because it never handles a real image.

\subsection{Provided tools}
Beyond the core physics, GOWC provides a set of
observer-facing tools, most driven by the same engine and live data. For
planning: a live weather map of all sites, object-visibility and observing-window
calculators, site comparison, a guided `Plan an Observation' wizard, and the
SNR/exposure calculator. For analysis: atmospheric analysis (seeing, PWV and
jet stream), historical reliability, and telescope efficiency. For transient and
event work: a Transient Follow-Up tab that lists active neutron-star X-ray
binaries with live outburst alerts from the MAXI monitor (this first version
covers X-ray binaries; other classes are flagged as forthcoming), and a Sky
Events section with trackers for meteor showers, eclipses, comets (Minor Planet
Center orbits), satellite passes (CelesTrak orbital elements), and near-Earth
asteroids (NASA's NeoWs service). A literature-search tool queries the NASA ADS
database. An observing assistant answers plain-language planning questions: a
large language model (Anthropic's Claude, through its API) only parses the
question into structured parameters and phrases the answer, while every number
in the answer comes from the deterministic engine; the assistant is rate-limited
and declines when the engine cannot answer. Registered users can save sites and
log observing sessions.

\subsection{Relation to existing tools}
The individual computations GOWC performs
are available elsewhere, but in separate tools: target visibility and airmass in
Astroplan \citep{astroplan}; sky brightness, extinction and exposure times in the
ESO exposure-time calculators and \textsc{skycalc} \citep{skycalc};
instrument-specific SNR in per-facility calculators such as ING \textsc{signal}
\citep{signal}. What GOWC adds is their combination in one place. Visibility,
live atmospheric conditions and photometric SNR are computed from one shared
physics module across a global catalogue of $\sim$2{,}600 sites rather than a
single facility, with each output traceable to a published model or stated
calibration and, where possible, tested against a site monitor. It is the breadth
(many sites, the whole planning chain) rather than the depth at any one facility
that distinguishes it.

\subsection{Installation, usage and reproducibility}
GOWC runs as a hosted web
application (\texttt{gowcastroclimate.com}); the physics engine is a standalone
Python package (Python $\ge$3.10; dependencies \texttt{numpy}, \texttt{pandas},
\texttt{ephem}, \texttt{requests}) installable from the
repository with \texttt{pip install .}\ and usable without the web layer. The
core entry points are direct function calls. For example,
\texttt{rank\_sites(ra\_deg, dec\_deg, date\_utc)} ranks a curated set of eight
well-placed observatories for a target by airmass, altitude during astronomical
darkness and current conditions, and \texttt{get\_full\_atmospheric\_analysis(record)}
returns the seeing, PWV and jet-stream diagnostics for a site from its weather
record. Every number in Tables~\ref{tab:val}--\ref{tab:multiyear} and~\ref{tab:snrgrid} and Figs~\ref{fig:pwv}--\ref{fig:ranking} (Fig.~\ref{fig:arch} is a
schematic; Table~\ref{tab:instr} lists the constants in \texttt{instruments.py}) is produced by
a re-runnable script in \texttt{tests/}: \texttt{validate\_snr\_etc.py} and
\texttt{snr\_grid\_compare.py} (Tables~\ref{tab:val} and~\ref{tab:snrgrid}, Appendix~\ref{app:snrgrid}),
\texttt{dimm\_mass\_decomposition.py} (Table~\ref{tab:decomp}), \texttt{site\_bias.py} (Table~\ref{tab:bias}),
\texttt{hindcast\_multiyear.py} (calibration constants, Table~\ref{tab:multiyear}, baselines),
\texttt{hindcast\_extras.py} (per-year and out-of-window splits, partial
correlations, poor-night skill), \texttt{validate\_pwv\_lhatpro.py} (Section~\ref{sec:pwv}), \texttt{pwv\_terrain\_gap.py} (Section~\ref{sec:pwvmodel}),
\texttt{validate\_airmass.py} and \texttt{validate\_extinction.py} (Section~\ref{sec:airmassval}), and
\texttt{make\_paper\_figures.py} / \texttt{make\_arch\_figure.py} (figures). This
paper describes GOWC v1.0.3 (MIT-licensed); the exact version is archived at the
Zenodo DOI given in the Data Availability section. Bugs and feature requests can be reported through the repository's GitHub issue tracker (see Data Availability).

\section{Physical Models}\label{sec:models}

All quantities below are computed from the cited formalism or the stated
calibrations. Constants are quoted as implemented.

\subsection{Seeing}
\label{sec:seeing}
Seeing (FWHM, arcsec, at $\lambda=500$\,nm) combines in quadrature a
free-atmosphere term, computed from the Tatarski $C_n^2$ formulation
\citep{tatarski} as re-derived by \citet{basu}, and an empirical
ground-layer term.

\emph{Free atmosphere.} For each layer between adjacent pressure levels above the
site (850, 700, 600 and 500\,hPa), the potential-temperature gradient
$\Gamma=d\theta/dz$, with $\theta=T(1000/P)^{R/c_p}$, gives the refractive-index
structure constant
\begin{equation}
C_n^2=\Big(\frac{79\times10^{-6}P}{T^2}\Big)^{2}\,
\frac{c}{\mathrm{Pr}_t}\,L_0^{4/3}\Big(\frac{T}{\theta}\Big)^{2}\Gamma^2 ,
\label{eq:cn2}
\end{equation}
evaluated with each layer's mean $T$, $P$ and $\theta$, where $c=2.8$ is the
Tatarski coefficient and $\mathrm{Pr}_t=1$ the turbulent Prandtl number for
stable night-time stratification. The factor $(T/\theta)^2$ converts the
potential-temperature fluctuations that the gradient argument yields into the
actual-temperature fluctuations that refract light \citep{basu}; it falls from
0.91 at 850\,hPa to 0.67 at 500\,hPa, a $\sim$14~per~cent effect on seeing. The outer
scale follows the tropospheric model of \citet{dewan} \citep[see also][]{trinquet},
\begin{equation}
L_0^{4/3}=0.1^{4/3}\times10^{\,1.64+42S},
\label{eq:dewan}
\end{equation}
with $S$ the wind shear across the layer (s$^{-1}$, from winds in m\,s$^{-1}$),
capped at $0.02$\,s$^{-1}$ (reached in 0.01~per~cent of Paranal layers) and set to
$3\times10^{-3}$\,s$^{-1}$ where no shear is available. Because $S$ sits in an
exponent, $\partial\ln\varepsilon/\partial S\approx58$\,s: a shear error of
$0.001$\,s$^{-1}$ is a $\sim$6~per~cent seeing error, which is why the wind units
matter (the median Paranal layer shear is $2.2\times10^{-3}$\,s$^{-1}$; winds
read as km\,h$^{-1}$ inflate it 3.6-fold). Each layer is weighted by its
geometric depth and the layers are summed. The lowest bracket is interpolated to
the telescope elevation, so levels below the site, including the Atacama marine
inversion near 1000--1500\,m, are excluded; a site below 850\,hPa
($\sim$1500\,m) integrates from 850\,hPa and leaves the slab beneath to the
ground-layer term. The Fried parameter \citep{fried} and the free-atmosphere
seeing are then
\begin{equation}
r_0=\Big(0.423\,k^2 X\,C\!\int\! C_n^2\,dh\Big)^{-3/5},\qquad
\varepsilon_{\text{free}}=0.98\,\lambda/r_0 ,
\label{eq:r0}
\end{equation}
with $k=2\pi/\lambda$, the airmass $X$ inside the bracket (slant path), and $C$
a single calibration constant.

\emph{Calibration.} $C$ is set at Paranal so that the median of
$\varepsilon_{\text{free}}$ over the 915 hours matched to ESO MASS measurements
(2026-05-29 to 08-29) equals the MASS median over the same hours ($0.45$~arcsec),
giving $C=2.03$. Because $c$ and $C$ multiply the same integral, only
$c_{\text{eff}}=cC=5.7$ is identifiable. This is nearly twice the upper end of the
published range of 2.8--3.2 \citep{basu}. We read this as a measurement: four pressure levels resolve
only broad gradients, not the thin sheets where optical turbulence concentrates,
and the calibration absorbs the factor of about two in $\int C_n^2\,dh$ that the
coarse profile misses. The same constant serves the live Forecast API and the
Historical Forecast API used for validation; on 48 Paranal readings stored by the
live site since 2026-08-25 the two give free-atmosphere medians of $0.50$~arcsec and
$0.49$~arcsec (rank correlation 0.99). Two caveats apply. MASS senses turbulence from
$\sim$500\,m above the site while the integral starts at the telescope, so the
slab up to the 700\,hPa level ($\sim$3150\,m at Paranal) is counted in
$\varepsilon_{\text{free}}$ but also in the DIMM$-$MASS ground layer of
Section~\ref{sec:dimm}; the calibration absorbs this overlap. Where the 700/600\,hPa
levels are missing, a two-level fallback uses the 850$\to$500\,hPa gradient over
an effective thickness calibrated the same way (4900\,m).

\emph{Ground layer and total.} The pressure levels do not resolve the ground
layer, so it is a calibrated surface parametrization of our own, combined with
the free atmosphere in quadrature (independent turbulent layers):
\begin{equation}
\varepsilon_{\text{bl}}=\varepsilon_0\big(1+(w/8)^2\big)^{0.3}
e^{-h/10\,000}X^{0.6},\qquad
\varepsilon=\sqrt{\varepsilon_{\text{free}}^2+\varepsilon_{\text{bl}}^2},
\label{eq:bl}
\end{equation}
with surface wind $w$ (m\,s$^{-1}$), site altitude $h$ (m), and
$\varepsilon_0=0.77$~arcsec set so that Paranal at its median surface wind
(2.8\,m\,s$^{-1}$) reproduces the measured ground-layer median of $0.61$~arcsec
(Section~\ref{sec:dimm}). The 10\,000\,m scale height is chosen rather than fitted,
since one site cannot constrain it, and makes the term nearly site-invariant
(factor 1.00 at sea level, 0.64 at 4500\,m), consistent with a ground layer set by
local topography and heating. A humidity factor reduced ranking skill and is not
included. Where a site has no usable profile (and on the historical-reliability
page, whose archive stores surface fields only), $\varepsilon_{\text{free}}$ is
replaced by the calibrated median $0.45\,\mathrm{arcsec}\,X^{0.6}$, and the result is indicative
only.

\subsection{Airmass and extinction}
\label{sec:airmass}
A single airmass definition, the interpolative formula of \citet{pickering}, is
used throughout for target visibility, site ranking, extinction, scintillation and
the SNR path: $X = 1/\sin\!\big(a + 244/(165+47a^{1.1})\big)$, with $a$ the
altitude in degrees. The one exception is the moonlit-sky model of
Section~\ref{sec:sky}, which uses \citeauthor{ks1991}'s own airmass expression as
part of that published model. Extinction follows the Bouguer law $\Delta m = k\,X$,
with transmission $10^{-0.4\,kX}$. Band coefficients are representative values for the Roque de los Muchachos
Observatory (Appendix~\ref{app:constants}; cf.\ \citealt{king}), scaled to the
site altitude by a single exponential,
$k(h) = k_{\text{ref}}\,e^{-(h - h_{\text{ref}})/H_k}$, with
$h_{\text{ref}}=2000$\,m and $H_k = 8000$\,m (close to the Rayleigh scale height
of 7.996\,km adopted by \citealt{king}); the scale factor is floored at $0.70$, so
sites above $\sim$4850\,m take the high-site limit. A single scale height is a
deliberate simplification: the Rayleigh component scales with pressure
($H\approx8$\,km) while the aerosol component falls far faster
($H\approx1$--2\,km), and the aerosol term, which is the most variable component
and can dominate $k$ at low sites, is the larger source of residual error. When a
target colour is supplied, the secondary term $k_2(B\!-\!V)$ is added to the
extinction coefficient, and the instrumental-to-standard colour term
$m_{\text{std}} = m_{\text{inst}} + \text{zp} + c_t\,(B\!-\!V)$ is computed and
reported alongside the result without changing the SNR
(Appendix~\ref{app:constants}). The current interface does not yet pass a colour,
so by default the single-band $k$ is used.

\subsection{Precipitable water vapour}
\label{sec:pwvmodel}
PWV uses the forecast model's total-column water vapour (kg\,m$^{-2}$, numerically
mm), which the model integrates over its full vertical resolution and is
therefore more reliable than integrating the two or three humidity levels GOWC
stores. Open-Meteo returns this column already adjusted, by statistical
downscaling, towards the height of a 90\,m digital elevation model at the
requested coordinates, and reports that height, $h_{\text{DEM}}$, with each query
\citep{openmeteo}. GOWC removes only the residual water between that height and
the telescope, $\mathrm{PWV}(h_{\text{site}}) = \mathrm{PWV}_{\text{OM}}\,
e^{-(h_{\text{site}}-h_{\text{DEM}})/H_w}$ with $H_w \approx 2000$\,m, applied only
when the site lies above $h_{\text{DEM}}$. Because GOWC takes its site altitudes
from the same elevation service, this residual correction is zero for
98.5~per~cent of catalogued sites, and the reported PWV is then the Open-Meteo
column itself. The exceptions (39 sites) are mostly co-located stations that share one weather
request, plus nine Maunakea telescopes that share a single summit altitude; the
largest gap is 125\,m, a 6~per~cent correction. (Scaling
from sea level instead, $e^{-h_{\text{site}}/H_w}$, removes water that the
returned column no longer contains; the LHATPRO comparison in
Section~\ref{sec:pwv} caught this.) Where the column product is unavailable, PWV
falls back to a layered integral of specific humidity, computed from the
relative humidity with the Magnus relation \citep{magnus}, over the surface, 850
and 500\,hPa levels.

\subsection{Sky brightness}
\label{sec:sky}
Moonlit sky brightness uses the \citet{ks1991} model. A moonless $V$-band zenith
baseline of $21.8$\,mag\,arcsec$^{-2}$, typical of a dark site, is converted to
the observing band with the mean dark-time colours of the night sky at Paranal
\citep[$U\!-\!V=+0.64$, $B\!-\!V=+0.96$, $R\!-\!V=-0.78$, $I\!-\!V=-2.06$;][]{patat2008}
and scaled to the target's line of sight as $B_{\text{zen}}\,10^{-0.4k(X-1)}X$
(more airglow path, partly offset by extinction), with that band's extinction
coefficient $k$ (Section~\ref{sec:airmass}). The scattered moonlight is added in
flux as a function of the lunar phase angle $\alpha$, the Moon's and target's
airmasses, and, strongly, the Moon--target separation, using the site's own
$k_V$. The phase angle follows from the illuminated fraction $f$ as
$\alpha=\arccos(2f-1)$; a linear mapping would be $\sim$0.4\,mag wrong in the
gibbous phases. The moonlight term is the model's $V$-band value in every band,
a simplification of a model calibrated in $V$ alone, so in bright time the sky
outside $V$ is less certain than in dark time. Narrowband H$\alpha$ and
[O\,\textsc{iii}] take the $R$- and $V$-band sky, the bands they use for
zero-point and extinction. Every page evaluates the model with the real Moon and
target positions for the chosen time and site; the planning wizard also offers
what-if Moon phases, placed at an assumed $90^\circ$ from the target.

\subsection{Signal-to-noise ratio}
\label{sec:snr}
The SNR calculator implements the standard CCD equation \citep{howell2006}
with scintillation:
\begin{equation}
\mathrm{SNR} = \frac{S}{\sqrt{S + B_{\mathrm{sky}} + D_c
+ N_{\mathrm{read}}^2 n_{\mathrm{pix}} + \sigma_{\mathrm{scint}}^2}},
\label{eq:snr}
\end{equation}
where $S$ is the source signal and $B_{\mathrm{sky}}$ and $D_c$ the sky and
dark-current signals summed over the $n_{\mathrm{pix}}$ pixels of the aperture (a
1-FWHM radius), all in electrons, and $N_{\mathrm{read}}$ is the per-pixel read
noise ($D_c$ is distinct from the telescope diameter $D$ below and from the DIMM
seeing of Section~\ref{sec:dimm}). Source flux follows from per-band Vega zero-points
\citep{bessell} and the instrument's throughput and quantum efficiency, or
directly from a published instrument zero-point where one exists
(Appendix~\ref{app:constants}). Scintillation follows the modified Young approximation of \citet[eq.~7]{osborn},
$\sigma_Y^2 = 10^{-5}C_Y^2 D^{-4/3}t^{-1}X^3 e^{-2h/H}$, with $D$ in m, $t$ in s
and $h$ in m, $X^3$ in place of their $(\cos\gamma)^{-3}$, $H=8000$\,m, and
$C_Y\approx1.5$, the mean of their measured site values (their table~1);
$\sigma_Y$ is a fractional fluctuation, so it enters the noise budget as
$\sigma_{\text{scint}} = \sigma_Y\,S$. The sky is attenuated by the same extinction as the source. Strictly,
a sky brightness measured from the ground already includes the atmosphere, so
this understates the sky by a factor $10^{-0.4kX}$ ($\sim$11~per~cent at $X=1$); it is the
\textsc{signal} convention, and the faint-end benchmark confirms it (removing the
attenuation moves the sky-limited ACAM points from $-0.2$~per~cent to $-5$ to $-7$~per~cent
against \textsc{signal}). The moonlit-sky model of Section~\ref{sec:sky} also returns a
ground-level brightness, so in the planner the same factor applies and the SNR of
faint, sky-limited targets is optimistic by $\sim$6~per~cent at $X=1$ and $\sim$9~per~cent at
$X=1.5$ in $V$.

The exposure time for a target SNR solves
$S^2t^2 - \mathrm{SNR}^2(S+B+D_c+KS^2)\,t - \mathrm{SNR}^2 R = 0$, where $S$, $B$
and $D_c$ are now rates (e$^-$\,s$^{-1}$), $R$ is the total read-noise variance,
and $K=\sigma_Y^2t$ is constant because $\sigma_Y^2\propto t^{-1}$: scintillation,
like shot noise, grows linearly in $t$, so the solution is exact. For INT/WFC at
zenith (1.0$''$ seeing, sky 21.5\,mag\,arcsec$^{-2}$, $k=0.12$), a $V=20$ star
reaches SNR\,100 in 118\,s, while a $V=8$ star reaches SNR\,30 in 3\,ms, far below
a practical shutter time; the planner flags such cases together with saturation
(assuming a 100\,000\,e$^-$ full well where the detector's is unknown). Instrument
constants (aperture, obstruction, pixel scale, read noise, full well, throughput,
QE, and an empirical/theoretical efficiency) come from a small cited catalogue
(Appendix~\ref{app:constants}). The magnitude uncertainty is reported as
$\sigma_m = 1.0857/\mathrm{SNR}$.

\section{Validation}\label{sec:validation}

Each quantity is tested against the best available independent reference:
airmass against the \citeauthor{kasten} formula, extinction against published
site coefficients, the SNR against a professional exposure-time calculator, and
water vapour and seeing against site-monitor measurements. Cloud cover and sky brightness
are not independently tested (Section~\ref{sec:cloud}, Section~\ref{sec:skyval}).

\subsection{Airmass and extinction}
\label{sec:airmassval}
Airmass is deterministic geometry, checked against the approximation formula of
\citet{kasten}: GOWC's Pickering values agree to better than 0.3~per~cent at and
above $5^\circ$ altitude (largest 0.27~per~cent, at $5^\circ$), and the two forms
diverge to $\sim$1.6~per~cent at $0.1^\circ$, far below any observing limit.
Extinction coefficients are \emph{set} to representative values for the Roque de
los Muchachos Observatory (Section~\ref{sec:airmass}), so agreement at La Palma is
by construction (to 0.002\,mag\,airmass$^{-1}$). Scaled to Paranal, they agree
with the median of ESO's FORS1 quality-control broadband coefficients for
2004--2009 \citep{eso_fors1_qc} to within 0.003\,mag\,airmass$^{-1}$ in $U$, $R$
and $I$ and 0.012 in $V$, and are 0.05 low in $B$; in every band the difference
lies well inside the period-to-period scatter (e.g.\ 0.08--0.22 in $V$). This is
a weak test of the altitude scaling: the 275\,m step from La Palma to Paranal
changes the coefficient by only 3.4~per~cent, so the agreement mainly re-tests
the adopted values, and the extinction scale height is essentially unconstrained
here.

\subsection{Precipitable water vapour against LHATPRO}
\label{sec:pwv}
\begin{figure}[htbp]
\centering
\includegraphics[width=0.6\textwidth]{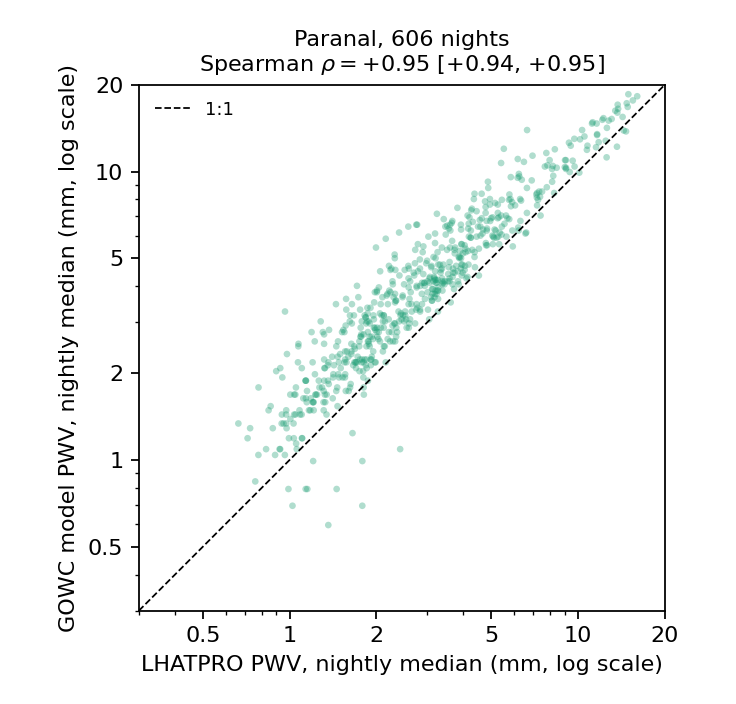}
\caption{GOWC PWV (the Open-Meteo total column at the site,
Section~\ref{sec:pwvmodel}) against the ESO LHATPRO radiometer at Paranal \citep{eso_asm},
nightly medians over 606 nights. The rank correlation is the strongest in the
paper ($\rho=+0.95$); almost all points sit above the 1:1 line, reflecting the
$+43$~per~cent median wet bias of the forecast column.}
\Description[Scatter plot of model against measured water vapour at Paranal.]{Log-log scatter plot of nightly GOWC model precipitable water vapour against LHATPRO measured water vapour at Paranal for 606 nights, spanning about 0.6 to 20 mm. The points form a tight band lying above the dashed one-to-one line, showing a strong correlation (Spearman rho 0.95) and a model that reads too wet.}%
\label{fig:pwv}
\end{figure}

GOWC's model column, corrected to the site (Section~\ref{sec:pwvmodel}), is compared with
the ESO LHATPRO radiometer at Paranal over 2025-01 to 2026-08, as nightly medians
with nights keyed noon to noon and the model sampled only at the hours LHATPRO
measured ($N=606$). The night-to-night agreement is excellent: Spearman
$\rho=+0.945$ ($[+0.94,+0.95]$), Pearson $r=+0.96$ (Fig.~\ref{fig:pwv}). The
comparison also diagnosed an error in an earlier altitude correction: the
Open-Meteo column is already referred to the elevation-model height at the site
($2617$\,m at Paranal), not to sea level, so scaling by $e^{-h_{\text{site}}/H_w}$
removed the sub-site water a second time and biased the column low by a factor of $3.7$
($2.6\times$ below the measurement), increasingly so with altitude across the
whole catalogue. With the correction the model median is $3.9$\,mm against a
measured $2.7$\,mm, a $+43$~per~cent wet bias (median per-night ratio $1.34$). The bias
is only mildly larger on the driest nights, which matter most for the infrared:
$1.44\times$ for nights below 1\,mm (28 nights) against $1.35\times$ above 4\,mm,
although a few dry nights scatter below the 1:1 line. GOWC therefore reports PWV
as a validated quantity for ranking ($\rho=0.95$) with a stated absolute bias, and
displays the LHATPRO measurement where it exists.

Two caveats bound this test. First, at Paranal the elevation-model height
($2617$\,m) is essentially the site altitude ($2635$\,m), so GOWC's residual
correction is $e^{-18/2000}=0.99$: LHATPRO validates the Open-Meteo column itself
and diagnoses the double count, but says nothing about the exponential form or
$H_w=2000$\,m, which matter little because the residual correction is zero at most
sites. Second, Open-Meteo's downscaling from the native model grid cell ($1885$\,m at
Paranal) to the elevation-model height is only partial: over the LHATPRO hours
the downscaled column is a median $1.26$ times drier than the raw grid-cell
column (requested with \texttt{elevation=nan}), whereas the exponential with
$H_w=2000$\,m over the same 732\,m would give $1.44$. Applying GOWC's correction
to the raw column instead, from the grid-cell height,
$\mathrm{PWV} = \mathrm{PWV}_{\text{raw}}\,e^{-(h_{\text{site}}-1885\,\text{m})/H_w}$,
reduces the wet bias from $+43$ to $+29$~per~cent but lowers $\rho$ from $0.945$
to $0.919$. Incomplete removal of the sub-site column therefore plausibly
explains part, though not all, of the wet bias; we keep the current method,
which ranks nights better. (Applying the grid-cell correction to the
already-downscaled column would remove the same slab twice.)

\subsection{Cloud cover: passed through, not validated}
\label{sec:cloud}
Cloud cover is the most decisive input for whether a night happens at all. GOWC
takes the provider's forecast cloud fraction $c$ and uses it in the headline
observing-quality score, a product of factors for cloud, humidity, dew-point
margin, wind, seeing, jet stream and precipitation, in which cloud enters as
$(1-c)^{1.5}$ so that even thin cover lowers the score sharply. The cloud forecast
itself is not independently validated here; it is passed through as supplied. A
satellite-cloud or all-sky-camera comparison is the obvious next test.

\subsection{Signal-to-noise ratio}
We compared the photometric SNR calculator with ING \textsc{signal}
\citep{signal}, the official ETC for the Isaac Newton Group telescopes, for two
instruments, using GOWC's own catalogue entries, which is the configuration an
observer actually gets. Both calculators were run for a point source with $m=20$
in $V$, a 100\,s exposure, $1.0$~arcsec seeing, airmass 1.0, extinction 0.12 and
a dark sky ($21.5$\,mag\,arcsec$^{-2}$). Table~\ref{tab:val} gives these anchor
points, and a 16-point grid in magnitude ($m_V=16$--$22$) and airmass
($X=1.0,1.5$) gives a median $|\Delta|$ of $1.9$~per~cent, a worst point of
$4.8$~per~cent, and a median signed ratio of $0.999$ (Appendix~\ref{app:snrgrid}).

The two instruments test different parts of the chain. GOWC's ACAM entry carries
\textsc{signal}'s own constants for the instrument (collecting area
$12.47$\,m$^2$, throughput $0.72\times0.71$, QE $0.85$, an 86\,nm bandwidth,
3.2\,e$^-$ read noise and 0.25\,arcsec pixels), so its agreement tests the
CCD-equation bookkeeping. (A published-zero-point configuration for ACAM,
$\mathrm{ZP1}_V=27.0$, gives an SNR 16~per~cent high against \textsc{signal},
whose counts imply $\approx26.7$, so the catalogue uses the computed path.) WFC
tests the throughput term: without it GOWC is high by a constant factor of about
1.4 in both source and sky counts, and because the factor is common to source and
sky it must be a throughput term. It matches \textsc{signal}'s
`empirical/theoretical' efficiency of $\sim$0.70 for WFC, which the catalogue
carries. The remaining WFC offset is a constant $\sim$6~per~cent in counts, or
$\sim$3~per~cent in SNR (Appendix~\ref{app:snrgrid}). Because both entries take
their constants from \textsc{signal}, this benchmark validates GOWC's
implementation of the CCD equation and its noise terms, not the independence of
its instrument constants; the other catalogued instruments (pt5m, VLT/FORS2,
Keck/LRIS, Gemini-N/GMOS-N) use literature values and are not benchmarked here.

\begin{table}[htbp]
\centering
\small
\setlength{\tabcolsep}{4pt}
\begin{tabular}{l r r r l}
\hline
\textbf{Instrument} & \textbf{GOWC SNR} & \textbf{SIGNAL SNR} &
\textbf{SNR ratio} & \textbf{object counts (GOWC/SIGNAL)} \\
\hline
WHT / ACAM / $V$ & 152.3 & 150.56 & 1.01 & 41{,}679 / 40{,}839 (2.1\%) \\
INT / WFC / $V$  &  92.1 &  89.20 & 1.03 & 15{,}303 / 14{,}425 (6.1\%) \\
\hline
\end{tabular}
\caption{GOWC vs.\ ING \textsc{signal} at the anchor point ($m_V=20$, 100\,s,
$1.0$~arcsec seeing, $X=1$, $k=0.12$, dark sky), from
\texttt{tests/validate\_snr\_etc.py}. A second \textsc{signal} run for ACAM, made
for the grid of Table~\ref{tab:snrgrid}, returned 152.66; since \textsc{signal} is
deterministic, some input must have differed between the two runs, and the 1.4~per~cent
difference indicates the precision of this hand-run comparison.}
\label{tab:val}
\end{table}

\subsection{Seeing: calibration and validation}
\label{sec:dimm}
\begin{figure}[htbp]
\centering
\includegraphics[width=0.6\textwidth]{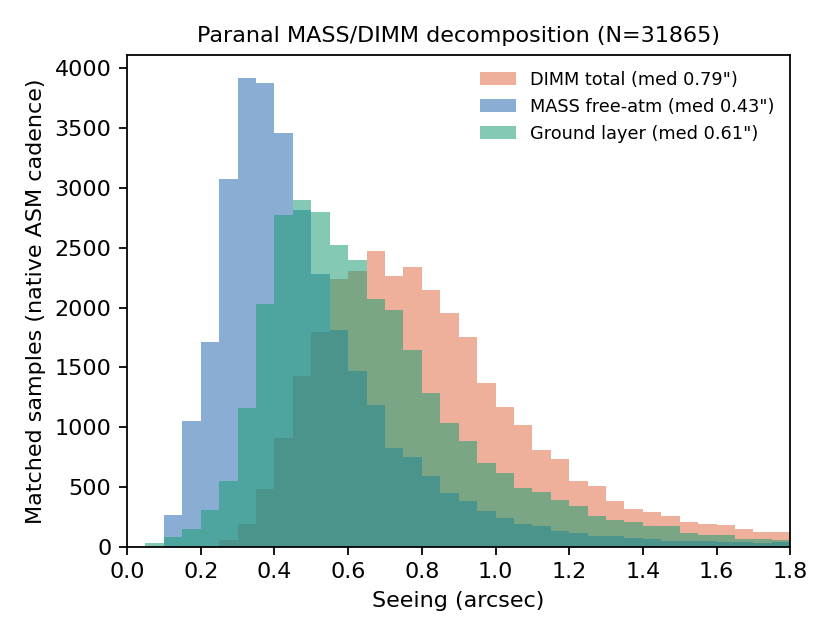}
\caption{The Paranal MASS/DIMM decomposition as distributions: DIMM total seeing,
MASS free-atmosphere seeing (above $\sim$500\,m), and the per-timestamp ground layer
$\sqrt{D^2-M^2}$, which carries the bulk of the total seeing and its variance.}
\Description[Histograms of measured seeing components at Paranal.]{Overlapping histograms of seeing at Paranal from 31,865 matched measurements. MASS free-atmosphere seeing peaks near 0.35 arcsec (median 0.43), the derived ground layer peaks near 0.5 arcsec (median 0.61), and DIMM total seeing is broadest, with median 0.79 arcsec and a tail beyond 1.5 arcsec.}%
\label{fig:decomp}
\end{figure}

The seeing model was developed against the ESO Astronomical Site Monitor at
Paranal \citep{eso_asm}, which runs both a Differential Image Motion Monitor (DIMM, total seeing)
and a Multi-Aperture Scintillation Sensor (MASS, free-atmosphere seeing above
$\sim$500\,m). Having both on the same timestamps gives the ground-layer
contribution in quadrature, computed per timestamp and then reduced to a median
(medians do not commute with quadrature subtraction), discarding the 0.6~per~cent of
pairs with $M\ge D$. The decomposition over 2026-05-29 to 08-29 ($N=31{,}865$
matched pairs) is given in Table~\ref{tab:decomp} and Fig.~\ref{fig:decomp}.

\begin{table}[htbp]
\centering
\begin{tabular}{l r l}
\hline
\textbf{Component (Paranal, measured)} & \textbf{Value} & \textbf{Reduction} \\
\hline
DIMM total seeing              & $0.79$~arcsec & median of $D$ \\
MASS free-atmosphere seeing    & $0.43$~arcsec & median of $M$ \\
Ground layer                   & $0.61$~arcsec & median of $\sqrt{D^2-M^2}$ (per pair) \\
Ground-layer variance fraction & 67\%     & median of $(D^2-M^2)/D^2$ (per pair) \\
ESO's own ground $C_n^2$ fraction & 65\%  & ESO MASS-DIMM product \\
\hline
\end{tabular}
\caption{Measured MASS/DIMM seeing decomposition at Paranal (native cadence). Rows
are not related by quadrature among their medians. The per-pair variance fraction
(67~per~cent) is corroborated by ESO's independently derived ground $C_n^2$ fraction (65~per~cent).
The window is southern winter. The calibration of Section~\ref{sec:seeing} uses the MASS
median over the hours matched to the forecast ($0.45$~arcsec), which differs slightly from
the native-cadence $0.43$~arcsec here.}
\label{tab:decomp}
\end{table}

\subsubsection{Calibration (Paranal)}
Two constants are calibrated at this site: $C$,
so that the free-atmosphere median matches MASS, and $\varepsilon_0$, so that the
ground layer matches the measured ground-layer median at median surface wind.
Agreement with the Paranal total is therefore largely a property of the
calibration and carries little model information. The informative tests are the held-out sites and the ranking hindcast, which we also repeat outside the calibration
window.

\subsubsection{Absolute scale at other sites}
Table~\ref{tab:bias} compares GOWC's
median seeing with a measured median at four sites. Even at Paranal the full
20-month median is $+7$~per~cent high, because the constants were fixed on the
southern-winter MASS hours and at median wind rather than over the whole
baseline.

\begin{table}[htbp]
\centering
\small
\setlength{\tabcolsep}{4pt}
\begin{tabular}{l r r l}
\hline
\textbf{Site} & \textbf{GOWC (arcsec)} & \textbf{Measured (arcsec)} & \textbf{Bias} \\
\hline
Paranal (calibration) & 0.80 & 0.75 (ASM DIMM, this work) & $+7\%$ \\
La Silla              & 0.85 & 0.79 (ASM DIMM, this work) & $+7\%$ \\
La Palma (ORM)        & 0.96 & 0.76 (DIMM mean) \citep{munoztunon} & $+26\%$ \\
Mauna Kea (13N)       & 0.64 & 0.75 (DIMM median) \citep{schoeck} & $-15\%$ \\
\hline
\end{tabular}
\caption{Absolute-bias check, median of nightly medians over 2025-01 to 2026-08:
at the DIMM-measured hours for Paranal and La Silla, and over local night hours
(20--06\,h solar time) for La Palma and Mauna Kea (\texttt{tests/site\_bias.py}).
The literature medians differ in epoch and instrument, and the La Palma value is
a mean.}
\label{tab:bias}
\end{table}

Over the three non-calibration sites the model's mean absolute bias is
16~per~cent, whereas a constant equal to Paranal's measured median $0.748$~arcsec scores 2.5~per~cent: the forecast's absolute
scale carries no site-to-site information that a constant does not, and at La
Palma it is 26~per~cent high. GOWC therefore does not claim to rank sites by forecast
seeing, and labels the absolute seeing it displays as indicative. Sites without a
verifiable published DIMM median (e.g.\ Hanle) are not included.

\subsubsection{Night-to-night ranking over 20 months}
\label{sec:ranking}
\begin{figure}[htbp]
\centering
\includegraphics[width=0.6\textwidth]{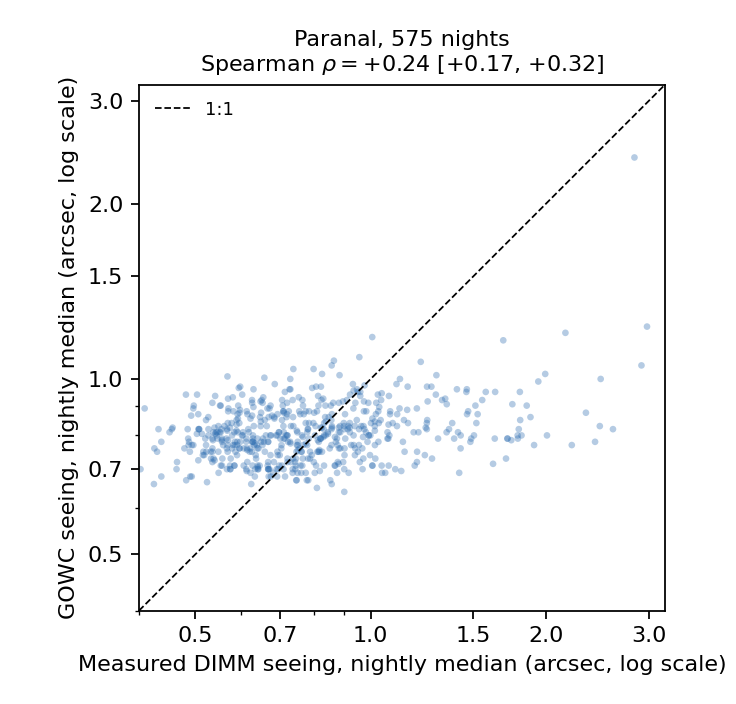}
\caption{Nightly-median GOWC seeing against measured DIMM at Paranal over 575 nights.
The rank correlation is weak but significant, and the model compresses the dynamic
range: its log spread is $0.32\times$ that of the DIMM and the OLS slope of
$\log$(model) on $\log$(DIMM) is $0.10$.}
\Description[Scatter plot of forecast against measured seeing at Paranal.]{Log-log scatter plot of nightly GOWC seeing against measured DIMM seeing at Paranal for 575 nights. Measured seeing spans about 0.4 to 3 arcsec while the model stays mostly between 0.65 and 1.0 arcsec, so the points form a nearly flat cloud crossing the one-to-one line, showing a weak correlation (Spearman rho 0.24) and a compressed range.}%
\label{fig:ranking}
\end{figure}

Bias and ranking are separate properties, so ranking is tested directly. GOWC's
seeing is hindcast through Open-Meteo's Historical Forecast API, which archives
pressure-level forecast fields (unlike Open-Meteo's ERA5 archive endpoint, which
does not serve pressure levels, or the live API, which keeps $\sim$12 days), with
every wind converted to m\,s$^{-1}$, and time-matched to the ESO ASM DIMM record
\citep{eso_asm} over 2025-01-01 to 2026-08-28. The ASM DIMM reports seeing
corrected to zenith, so the model is evaluated at $X=1$. DIMM seeing
autocorrelates over 1--3 hours, so each observing night, keyed noon to noon in
local time, is reduced to one median, and the model is sampled only at the hours
the DIMM measured. This archive stores each run's initial hours, i.e.\
near-analysis rather than a multi-day-ahead forecast, so every $\rho$ below
should be read as an upper bound on operational planning skill. Results are in
Table~\ref{tab:multiyear} and Fig.~\ref{fig:ranking}.

\begin{table}[htbp]
\centering
\begin{tabular}{l r r l}
\hline
\textbf{Predictor of nightly DIMM (or MASS)} & \textbf{$N$} & \textbf{Spearman $\rho$} & \textbf{95\% CI} \\
\hline
\multicolumn{4}{l}{\emph{Paranal}}\\
GOWC total seeing vs DIMM                & 575 & $+0.24$ & $[+0.17,\,+0.32]$ \\
\quad outside the calibration window     & 494 & $+0.25$ & $[+0.16,\,+0.33]$ \\
\quad deseasonalized (monthly medians)   & 575 & $+0.26$ & \\
GOWC $\varepsilon_{\text{free}}$ vs MASS & 82  & $+0.35$ & $[+0.14,\,+0.52]$ \\
Surface wind alone vs DIMM               & 575 & $+0.27$ & \\
Persistence (last night's DIMM)          & 559 & $+0.42$ & \\
\multicolumn{4}{l}{\emph{La Silla}}\\
GOWC total seeing vs DIMM                & 359 & $+0.16$ & $[+0.06,\,+0.26]$ \\
\quad deseasonalized                     & 359 & $+0.36$ & \\
Surface wind alone vs DIMM               & 359 & $+0.41$ & \\
Persistence                              & 321 & $+0.47$ & \\
\hline
\end{tabular}
\caption{Night-to-night ranking skill on nightly medians, from
\texttt{hindcast\_multiyear.py} and \texttt{hindcast\_extras.py}. Fisher-$z$ intervals;
a moving-block bootstrap agrees (Paranal $[+0.16,+0.33]$, La Silla $[+0.04,+0.30]$).
By year, Paranal gives $+0.27$ in 2025 ($N=352$) and $+0.19$ in 2026 ($N=222$), plus one
night keyed 2024-12-31 by the first UTC hours of 2025; La Silla gives $+0.40$ in 2025
($N=190$) and $-0.05$ in 2026 ($N=169$), so its full-baseline value is carried
entirely by 2025. A stricter $\ge$3-hours-per-night filter changes nothing (Paranal
$+0.24$, $N=570$).}
\label{tab:multiyear}
\end{table}

Five points summarize the hindcast. (i) Every interval at Paranal excludes zero,
so the model does carry night-to-night information, but it is weak:
$\rho^2\approx6$~per~cent of the rank variance, a debiased log scatter of
0.15\,dex, and a strongly compressed dynamic range. (ii) It is not an artefact of
the calibration window ($\rho=+0.25$ outside it) or of the seasons
(deseasonalized $\rho=+0.26$ at Paranal). At La Silla deseasonalizing raises
$\rho$ from $+0.16$ to $+0.36$: there the model's seasonal cycle runs against the
measured one and suppresses the raw score. (iii) The baselines an observer
already has do as well or better: surface wind alone scores $+0.27$ at Paranal
and $+0.41$ at La Silla, and persistence $+0.42$ and $+0.47$; the full model
beats neither. Controlling for persistence, it retains a partial $\rho$ of
$+0.22$ (Paranal) and $+0.18$ (La Silla). (iv) The free-atmosphere term carries
independent information. Its marginal skill against DIMM is modest ($+0.21$,
$+0.14$), but it tracks MASS ($\rho=+0.35$, a test immune to the calibration,
since $C$ is a monotonic scale on $\varepsilon_{\text{free}}$ and
$\varepsilon_0$ does not enter it), and controlling for surface wind it retains
a partial $\rho$ of $+0.28$ at Paranal and $+0.40$ at La Silla. (v) Part of the
weakness is therefore in the combination: surface wind and the free-atmosphere
term each carry information, but adding them in quadrature with weights
calibrated to Paranal's median conditions does not exploit it, and at La Silla
the combined forecast scores below its own ground-layer input ($+0.16$ against
$+0.41$; at a fixed site the ground-layer term is surface wind re-expressed).

Cast as an operational question, `will tonight be poor?' (DIMM $>1.0$~arcsec, a
21~per~cent base rate at Paranal), a model threshold set near the base rate
($0.88$~arcsec, flagging 22~per~cent of nights) catches 36~per~cent of poor
nights while flagging 19~per~cent of good ones. When it flags a night as poor it
is right 33~per~cent of the time, a $1.6\times$ enrichment over the base rate.
This is measured at Paranal, where GOWC displays the measured seeing instead; it
indicates what to expect at sites without a monitor.

The rest of the shortfall plausibly reflects the input resolution. Dedicated
mesoscale optical-turbulence forecasts that resolve the vertical turbulence
profile (e.g.\ the MOSE/Astro-Meso-NH system at Paranal \citep{masciadri})
achieve much higher skill, at far greater computational cost and for one site at
a time. GOWC drives the published Tatarski chain from four standard pressure
levels across a global catalogue; the $c_{\text{eff}}\approx5.7$ calibration and
the compressed dynamic range both indicate that those levels miss much of the
structure that sets the seeing.

This result does not affect the geometric part of site selection, namely which
observatory can observe an object tonight, for how long and at what airmass. That
part rests on the airmass and dark-time calculations validated in
Section~\ref{sec:airmassval} and on the adopted extinction of
Section~\ref{sec:airmass}; forecast seeing enters the observing-quality score
only as one factor among several. Where a public site monitor exists, GOWC also
displays the measured value directly: live DIMM seeing from the ESO Paranal ASM
\citep{eso_asm} and the Maunakea Weather Center \citep{mkwc_seeing}, and live
PWV from the Paranal LHATPRO radiometer.

\subsection{Sky brightness: modelled, not validated here}
\label{sec:skyval}
Moonlit sky brightness (Section~\ref{sec:sky}) is computed from the widely used
\citet{ks1991} model, with the dark-time sky colours of \citet{patat2008} and a
colourless moonlight term. We do not validate it independently here. It matters
because it sets the sky background in every SNR and exposure-time estimate. For a
sky-limited target, a sky that is $\Delta m$ too bright lowers the SNR by a
factor $10^{-0.2\Delta m}$ (about 13~per~cent for $0.3$\,mag), so any error
propagates directly into faint-target exposure times, while bright,
source-limited targets are barely affected. The natural next check is a grid
comparison against the ESO \textsc{skycalc} model \citep{skycalc} across band,
lunar phase and Moon--target separation (Section~\ref{sec:untested}).

\section{Discussion and Limitations}\label{sec:discussion}

Every physical model in GOWC is taken from the published literature except the
ground-layer term $\varepsilon_{\text{bl}}$ (Section~\ref{sec:seeing}), a
calibrated parametrization of our own; the observing-quality score
(Section~\ref{sec:cloud}), the adopted extinction coefficients and their
single-exponential altitude scaling (Section~\ref{sec:airmass}) are likewise our
own design choices. The outputs separate into three groups.

\emph{On solid ground:} airmass ($<0.3$~per~cent against the
\citeauthor{kasten} formula at and above $5^\circ$), the CCD SNR equation on the
two benchmarked instruments (median $1.9$~per~cent against \textsc{signal} over
16 points, Appendix~\ref{app:snrgrid}), the Bouguer extinction form with
representative coefficients (within the period-to-period scatter of ESO's
Paranal measurements, although the test mainly re-checks the adopted values),
and precipitable water vapour, which tracks LHATPRO night to night ($\rho=0.95$)
with a stated $+43$~per~cent wet bias.

\emph{Indicative:} forecast seeing. Its free-atmosphere term is physically
meaningful, since it tracks MASS and carries information independent of surface
wind, but the full forecast ranks nights only weakly ($\rho=+0.24$ at Paranal,
$+0.16$ at La Silla), does no better than surface wind or persistence, and has
no site-to-site skill in absolute scale. The calibration requires
$c_{\text{eff}}\approx5.7$, nearly twice the upper end of the published range,
which we interpret as the coarse vertical resolution missing roughly half of
$\int C_n^2\,dh$. The most direct improvement is to weight the free-atmosphere
and ground terms by a fit to the ranking rather than to Paranal's medians, since
the partial correlations show that both carry information the quadrature sum
wastes; transferability to a third site in a different climate regime is
currently assumed, not demonstrated. Section~\ref{sec:future} sets out these
steps.

\emph{Passed through or modelled only:} cloud cover (Section~\ref{sec:cloud})
and moonlit sky brightness (Section~\ref{sec:skyval}), whose moonlight term is
colourless. Night-to-night changes in aerosol and dust extinction are not
modelled, and four of the six catalogued instruments use literature constants
that are not benchmarked here.

\subsection{Forecast data and vertical resolution}
\label{sec:dimm-inv}
GOWC uses \emph{forecast} fields, not on-site measurements; the forecast sets a
floor on accuracy that no downstream model can improve, and real-time
transparency (thin cirrus, Saharan dust) is not captured. Each stored record is
the forecast hour nearest the fetch, not a night-integrated value. Even with four
pressure levels the profile cannot resolve the thin turbulent sheets that
dominate free-atmosphere seeing. At a high site the 850\,hPa level
($\sim$1500\,m) lies below the telescope and, in the Atacama, inside the marine
subsidence inversion; integrating only above the site removes that slab. A
corollary is that the number of usable layers falls with site altitude, because
the integral stops at 500\,hPa ($\sim$5850\,m at Paranal): Paranal has three
levels above it; a site above the 600\,hPa level ($\sim$4400\,m) keeps only
500\,hPa and integrates a single layer; a site above 500\,hPa would fall back to
the two-level form (no catalogued site does; the highest is $5369$\,m). Fetching
300 and 250\,hPa temperatures would add layers at every site and is a
straightforward improvement.

\subsection{Sky-brightness model}
The moonlight model (Section~\ref{sec:sky}) needs only
quantities GOWC already has, which is why it is used across the catalogue. More recent models are more accurate: \citet{jones2013} replace the empirical
fit with a physical scattering calculation, and \citet{winkler2022} revises the
simplified scattering model using photometry at SAAO. Both require site-specific
aerosol characterization and full sky-background components that a
forecast-driven planner covering thousands of sites cannot supply.

\subsection{Noise sources and simplifications in the SNR}
The SNR uses the standard CCD noise terms (source shot noise, sky, dark current,
read noise and scintillation), which dominate optical point-source photometry.
Thermal background (significant only beyond $\sim$2\,$\mu$m), flat-fielding residuals (a
$\sim$0.1--1~per~cent floor for very bright sources), fringing, cosmic rays and
charge-transfer inefficiency are omitted. Two simplifications also apply: in the
planner the modelled sky is attenuated a second time (Section~\ref{sec:snr}), making
faint-target SNR optimistic by $\sim$6--9~per~cent; and the narrowband H$\alpha$ and
[O\,\textsc{iii}] options treat the target as a continuum source, so their SNR is
only indicative for emission-line objects.

\section{Future Work}\label{sec:future}

\subsection{Seeing forecast}
A four-level global forecast cannot match a dedicated
single-site turbulence model such as MOSE \citep{masciadri}, but the hindcast shows
that the present model discards information it already has. Four steps address this:
\begin{enumerate}\setlength{\itemsep}{2pt}
\item \textbf{Statistical combination of the two terms.} Replace the fixed
quadrature sum, whose weights are set by Paranal's medians, with weights fitted
to the night-to-night ranking, which is standard statistical post-processing of
numerical forecasts. The partial correlations (Section~\ref{sec:dimm}) show that both
terms carry independent information. The fit will be tested out of sample:
trained on 2025 and scored on 2026, and trained at Paranal and scored at La
Silla, where the combined model currently scores below surface wind alone.
\item \textbf{Ground-layer stability.} The ground layer carries about two-thirds
of the seeing variance (Table~\ref{tab:decomp}) but is predicted from surface
wind alone. The provider also serves temperature and wind at 80, 120 and 180\,m
above ground; the near-surface lapse rate and shear they give measure the
stability that governs ground-layer turbulence, and are potentially the largest
single gain.
\item \textbf{More pressure levels.} The provider serves many more levels than
the four used here (e.g.\ 800, 400, 300, 250 and 200\,hPa). Adding them refines
the free-atmosphere integral, includes the jet-stream layer directly, and tests
whether finer vertical sampling brings $c_{\text{eff}}$ back towards the
published range. A repeat of the hindcast on ERA5 \citep{hersbach2020}, with 37 pressure levels and a
homogeneous reanalysis, would test the same question independently.
\item \textbf{Blending with persistence at monitored sites.} Persistence scores
higher than the model ($+0.42$ and $+0.47$) but the model retains independent
information ($\text{partial}\ \rho=+0.22$ and $+0.18$), so a blend of last night's
measured seeing and the forecast should outperform both where a monitor exists.
\end{enumerate}
Two evaluation steps accompany these: scoring the live forecasts, which GOWC
already logs twice daily for the DIMM-equipped sites, against the measured
seeing, to obtain the true planning skill at operational lead times rather than
the near-analysis upper bound reported here; and a third hindcast site in a
different climate regime (e.g.\ the ING RoboDIMM archive at La Palma) to test
transferability.

\subsection{Untested inputs}\label{sec:untested}
Compare the forecast cloud cover with satellite
cloud masks or all-sky-camera records; compare the moonlit sky against
\textsc{skycalc} \citep{skycalc} across band, lunar phase and Moon--target separation;
and test, at several sites with public radiometers, whether referring the PWV
column to the native model grid-cell height reduces the wet bias without degrading
the night-to-night ranking.

\subsection{Extinction and photometry}
Add forecast aerosol optical depth (e.g.\
from CAMS) to the extinction coefficient; benchmark the remaining catalogued
instruments against their own facility ETCs; treat the modelled sky as a
ground-level quantity; accept a target colour; and model narrowband SNR from
line fluxes.

\subsection{The reverse half of the pipeline}
GOWC implements the forward half
of the photometric pipeline. Extending it to the reverse half would reuse the
calibration mathematics already in the engine: reading raw FITS frames
(via \texttt{astropy}); bias, dark and flat-field correction with cosmic-ray
rejection; source detection, centroiding and aperture photometry with annulus
sky estimation (via \texttt{photutils}); calibration through the existing
extinction, zero-point and colour-term machinery; and differential photometry
and light curves.

A standalone web front-end to replace the current dashboard is in development;
it will use the same physics engine and validation scripts described here.

\section{Conclusion}

GOWC combines visibility, atmospheric conditions and instrument performance in a
single open planning tool covering $2{,}642$ catalogued observatory sites. Its SNR
calculator agrees with the ING \textsc{signal} ETC to a median $1.9$~per~cent across a
magnitude--airmass grid on two benchmarked instruments, and its precipitable water
vapour tracks the Paranal LHATPRO radiometer ($\rho=+0.95$) with a $+43$~per~cent wet
bias. Its seeing forecast, driven by four pressure levels, needs an effective
Tatarski coefficient nearly twice the upper end of the published range and ranks
nights only weakly ($\rho=+0.24$ at Paranal over 20 months), no better than
surface wind or persistence, although its free-atmosphere term tracks MASS and
carries information the surface wind does not. GOWC therefore treats forecast
seeing as indicative, shows measured values wherever a site monitor publishes
them, and states for each output whether it is measured, calibrated, modelled or
passed through. Combining the free-atmosphere and ground-layer terms by a fit to
the observed ranking, and adding near-surface stability and more pressure levels,
are the clearest routes to a more skilful seeing forecast (Section~\ref{sec:future}).

\section*{Acknowledgements}
This work relies on public data and services provided by others. The seeing and
precipitable-water validation uses the European Southern Observatory Astronomical
Site Monitor public archive (DIMM, MASS and LHATPRO data), and live seeing is also
taken from the Maunakea Weather Center. Meteorological inputs and site elevations
are from Open-Meteo, and the site list from the IAU Minor Planet Center. The SNR
validation compares against the \textsc{signal} exposure-time calculator provided
by the Isaac Newton Group of Telescopes. This research has made use of NASA's
Astrophysics Data System Bibliographic Services, and of MAXI data \citep{maxi} provided by
RIKEN, JAXA and the MAXI team; satellite orbital elements are from CelesTrak and
near-Earth-object data from NASA's NeoWs service. The observing assistant uses
Anthropic's Claude API for language parsing and phrasing only. The author used
Claude (Anthropic) to assist with reviewing the code and analysis and with editing
the manuscript; all changes were checked by the author, who takes full
responsibility for the content.

\emph{Software:} this work made use of NumPy \citep{harris2020}, pandas
\citep{pandas}, PyEphem \citep{pyephem}, Matplotlib \citep{hunter2007}, Requests
\citep{requests}, Streamlit \citep{streamlit}, Plotly \citep{plotly}, Folium
\citep{folium}, FastAPI \citep{fastapi}, psycopg2 \citep{psycopg2} with PostgreSQL
(hosted on Supabase), and Beautiful Soup \citep{bs4}.

\section*{Funding}
This research received no specific grant from any funding agency.

\section*{Conflict of Interest}
The author declares no conflict of interest.

\section*{Data Availability}
The source code, the physics module, and the re-runnable validation scripts
listed in Section~\ref{sec:overview} are released under the MIT licence at
\texttt{github.com/AhzamAhd/observatory-weather} and archived at Zenodo (v1.0.3,
DOI \href{https://doi.org/10.5281/zenodo.23541507}{\texttt{10.5281/zenodo.23541507}}). The archive also contains the cached
input files from which every number in this paper is computed: the Open-Meteo
historical-forecast profiles and the ESO ASM DIMM, MASS and LHATPRO extracts used in
the validation. Seeing and PWV validation data are from the public ESO
Astronomical Site Monitor archives for DIMM, MASS and LHATPRO \citep{eso_asm};
live seeing is also taken from the Maunakea Weather Center \citep{mkwc_seeing}.
Meteorological inputs are from Open-Meteo's
Forecast and Historical-Forecast APIs \citep{openmeteo} (surface and 850/700/600/500/250\,hPa
levels), site elevations from the Open-Meteo elevation service, and the site list
from the IAU Minor Planet Center observatory codes \citep{mpc_obscodes} , and the Paranal extinction comparison values from ESO's FORS1
quality-control pages \citep{eso_fors1_qc}. The \textsc{signal} values
used in the SNR benchmark are recorded in the benchmark scripts. No proprietary
data are used. The repository includes a \texttt{CITATION.cff}; please cite this
paper and the archived software when using GOWC. Bugs and feature requests can be
reported through the GitHub issue tracker at \texttt{github.com/AhzamAhd/observatory-weather/issues}.


\clearpage
\appendix
\section{Instrument and extinction constants}
\label{app:constants}
Table~\ref{tab:instr} lists the constants the SNR path (Section~\ref{sec:snr}) uses,
taken directly from the released \texttt{instruments.py}. `Eff.' is the
empirical/theoretical efficiency (\textsc{signal}'s term), set to $1.00$ where no
measured value is available. ZP1 is the zero-point expressed as the magnitude
giving $1\,\mathrm{e}^-\,\mathrm{s}^{-1}$ above the atmosphere; `comp.' marks
modes run on the computed photon-counting path (throughput\,$\times$\,QE).
`Obstr.' is the linear diameter ratio of the central obstruction.

\begin{table}[htbp]
\centering
\small
\setlength{\tabcolsep}{4pt}
\begin{tabular}{lccccccccc}
\hline
Facility/instr. & $D$ (m) & Obstr. & Pix (arcsec) & RN (e$^-$) & Full well & QE & Thr. & Eff. & ZP1 \\
\hline
INT/WFC & 2.54 & 0.30 & 0.333 & 3.5 & 135k & 0.80 & 0.76 & 0.70 & comp. \\
WHT/ACAM & 4.20 & 0.32 & 0.250 & 3.2 & 120k & 0.85 & 0.51 & 1.00 & comp. \\
pt5m/Sci.\ cam. & 0.50 & 0.35 & 0.280 & 8.0 & 40k & 0.60 & 0.55 & 1.00 & comp. \\
VLT/FORS2 & 8.20 & 0.14 & 0.250 & 2.9 & 200k & 0.90 & 0.70 & 1.00 & 28.0 ($V$) \\
Keck/LRIS & 10.00 & 0.14 & 0.135 & 4.0 & 100k & 0.88 & 0.55 & 1.00 & 28.7 ($V$) \\
Gemini-N/GMOS-N & 8.10 & 0.14 & 0.081 & 4.1 & 130k & 0.88 & 0.60 & 1.00 & 28.3 ($g'$) \\
\hline
\end{tabular}
\caption{Instrument constants used by the SNR calculator, from
\texttt{instruments.py}. Sources: INT/WFC and WHT/ACAM from the ING \textsc{signal}
ETC (ACAM: \citealt{benn}); the WHT obstruction is set so the clear collecting area
equals \textsc{signal}'s $12.47$\,m$^2$; pt5m \citep{hardy}; VLT/FORS2
\citep{appenzeller} and the ESO ETC; Keck/LRIS \citep{oke}; Gemini-N/GMOS-N
\citep{hook}, which has no $V$ filter, so its $g'$ zero-point is listed. QE and
Thr.\ apply only on the computed path. Some values are documented estimates where
a hard citation was unavailable.}
\label{tab:instr}
\end{table}

Extinction (Section~\ref{sec:airmass}) uses per-band reference coefficients
$k_{\text{ref}}$ tabulated at $h_{\text{ref}}=2000$\,m and scaled by
$k(h)=k_{\text{ref}}\,e^{-(h-h_{\text{ref}})/H_k}$, $H_k=8000$\,m, with the scale
factor floored at $0.70$. These give representative ORM values at La Palma (2360\,m; cf.\ \citealt{king}),
e.g.\ $k_V=0.125$ at $2000$\,m gives $0.120$ at La Palma and $0.116$ at Paranal. The reference set is $k_U=0.479$, $k_B=0.229$, $k_V=0.125$,
$k_R=0.094$, $k_I=0.052$, and $k_J=0.031$, $k_H=0.021$, $k_K=0.015$
(mag\,airmass$^{-1}$; the near-IR values are for transmission only). The
colour-dependent term $k=k_1+k_2(B\!-\!V)$ is applied when a target colour is
supplied; by default the single-band $k$ is used.

\section{SNR agreement across the observing regime}
\label{app:snrgrid}
Table~\ref{tab:snrgrid} compares GOWC, with its catalogue entries, against ING
\textsc{signal} on a grid in source magnitude ($m_V=16$--$22$, a factor
$\sim$250 in flux) and airmass ($X=1.0,1.5$; GOWC is evaluated at the altitude
whose Pickering airmass is exactly $X$), at 100\,s, $V$ band, dark sky
($21.50$\,mag\,arcsec$^{-2}$), extinction $0.12$ and $1.0$~arcsec seeing.
\textsc{signal} was run by hand; GOWC is recomputed by
\texttt{tests/snr\_grid\_compare.py}. Over the 16 points the median $|\Delta|$ is
$1.9$~per~cent, the worst is $4.8$~per~cent, and the median signed ratio is $0.999$.

The residuals are the sum of two effects, separated by re-running GOWC with its
scintillation term switched off (also in the script).

\emph{A constant throughput offset (INT/WFC).} With scintillation off, the WFC
residual is flat at $+3.0$ to $+3.6$~per~cent across all magnitudes and both airmasses.
GOWC's source and sky counts are both $\sim$6~per~cent above \textsc{signal}'s, and a
common factor $(1+\delta)$ on source and sky scales the SNR by $\sqrt{1+\delta}$
whenever read noise is negligible, whether the target is shot-noise-limited or
sky-limited, as here. A 6~per~cent count offset therefore appears as a
$\sim$3~per~cent SNR offset everywhere. ACAM, with
scintillation off, is flat at $-0.1$ to $-0.4$~per~cent.

\emph{Scintillation (both instruments).} The remaining structure grows with source
brightness and with airmass, the signature of the one noise term that scales with
the source counts and as $X^{1.5}$ in amplitude: at $X=1.5$ the ACAM residual runs
$-0.1,-0.3,-1.0,-4.8$~per~cent from $m_V=22$ to 16. For ACAM at $m_V=16$, GOWC's SNR falls
$5.7$~per~cent from $X=1.0$ to $1.5$ ($1256.7\to1185.0$) while \textsc{signal}'s falls
$2.7$~per~cent ($1280.0\to1245.1$); with scintillation off GOWC also falls $2.7$~per~cent. The
largest disagreement in the grid is therefore consistent with \textsc{signal} not
including scintillation for these settings. The same excess explains the one
small WFC point, $m_V=16$, $X=1.5$ ($-0.4$~per~cent), where scintillation cancels the
$+3$~per~cent throughput offset, and the smaller WFC offset at the bright end at $X=1.0$
($+1.9$~per~cent against $+3.0$~per~cent without scintillation).

\begin{table}[htbp]
\centering
\small
\begin{tabular}{lccccc}
\hline
Instrument & $m_V$ & $X$ & GOWC SNR & \textsc{signal} SNR & $\Delta$ \\
\hline
WHT/ACAM & 16 & 1.0 & 1256.7 & 1280.0 & $-1.8\%$ \\
WHT/ACAM & 18 & 1.0 & 482.2 & 484.8 & $-0.5\%$ \\
WHT/ACAM & 20 & 1.0 & 152.3 & 152.7 & $-0.2\%$ \\
WHT/ACAM & 22 & 1.0 & 33.10 & 33.15 & $-0.2\%$ \\
WHT/ACAM & 16 & 1.5 & 1185.0 & 1245.1 & $-4.8\%$ \\
WHT/ACAM & 18 & 1.5 & 466.8 & 471.6 & $-1.0\%$ \\
WHT/ACAM & 20 & 1.5 & 148.0 & 148.5 & $-0.3\%$ \\
WHT/ACAM & 22 & 1.5 & 32.20 & 32.23 & $-0.1\%$ \\
INT/WFC & 16 & 1.0 & 764.5 & 750.2 & $+1.9\%$ \\
INT/WFC & 18 & 1.0 & 292.2 & 283.9 & $+2.9\%$ \\
INT/WFC & 20 & 1.0 & 92.1 & 89.2 & $+3.3\%$ \\
INT/WFC & 22 & 1.0 & 20.00 & 19.31 & $+3.6\%$ \\
INT/WFC & 16 & 1.5 & 727.0 & 729.7 & $-0.4\%$ \\
INT/WFC & 18 & 1.5 & 283.3 & 276.2 & $+2.6\%$ \\
INT/WFC & 20 & 1.5 & 89.6 & 86.7 & $+3.3\%$ \\
INT/WFC & 22 & 1.5 & 19.40 & 18.77 & $+3.4\%$ \\
\hline
\end{tabular}
\caption{GOWC vs ING \textsc{signal} across the magnitude--airmass grid ($V$,
$100$\,s, sky $21.50$\,mag\,arcsec$^{-2}$, extinction $0.12$, seeing $1.0$~arcsec).
$\Delta$ is $(\mathrm{GOWC}-\textsc{signal})/\textsc{signal}$.}
\label{tab:snrgrid}
\end{table}


\end{document}